\documentclass[twocolumn]{aastex701}

\begin{document}
\title{Modeling the Optical Colors of Galactic Cirrus Clouds in the Stripe 82 Region}


\author[orcid=0000-0001-9561-8134]{Kwang-Il Seon}
\affiliation{Korea Astronomy and Space Science Institute, 776 Daedeokdae-ro, Yuseong-gu, Daejeon, 34055, Republic of Korea}
\affiliation{Department of Astronomy and Space Science, University of Science and Technology, Korea, 217 Gajeong-ro, Yuseong-gu, Daejeon, 34113, Republic of Korea}
\email[show]{kiseon@kasi.re.kr}

\author[orcid=0000-0002-9434-5936]{Jongwan Ko}
\affiliation{Korea Astronomy and Space Science Institute, 776 Daedeokdae-ro, Yuseong-gu, Daejeon, 34055, Republic of Korea}
\affiliation{Department of Astronomy and Space Science, University of Science and Technology, Korea, 217 Gajeong-ro, Yuseong-gu, Daejeon, 34113, Republic of Korea}
\email{jwko@kasi.re.kr}

\author[orcid=0000-0002-7762-7712]{Woowon Byun}
\affiliation{Korea Astronomy and Space Science Institute, 776 Daedeokdae-ro, Yuseong-gu, Daejeon, 34055, Republic of Korea}
\email{wbyeon@kasi.re.kr}

\author[orcid=0000-0002-6810-1778]{Jaehyun Lee}
\affiliation{Korea Astronomy and Space Science Institute, 776 Daedeokdae-ro, Yuseong-gu, Daejeon, 34055, Republic of Korea}
\email{jaehyun@kasi.re.kr}

\author[orcid=0000-0003-3574-1784]{Young-Soo Jo}
\affiliation{Korea Astronomy and Space Science Institute, 776 Daedeokdae-ro, Yuseong-gu, Daejeon, 34055, Republic of Korea}
\email{stspeak@kasi.re.kr}

\begin{abstract}
Observations have shown that the optical colors of Galactic cirrus clouds differ significantly from those of extragalactic sources; thus, they can be used to distinguish Galactic cirrus from extragalactic low surface brightness (LSB) features.
To understand these properties, we calculate radiative transfer models in dust clouds, where photons are incident from the ambient interstellar medium (ISM).
Dust clouds are modeled to mimic a turbulent medium using a fractional Brownian motion algorithm, resulting in a lognormal density distribution and a power-law power spectral density that are appropriate for the ISM.
The results are compared with optical observations of cirrus clouds in the Stripe 82 region.
The observed color--color ($g-r$, $r-i$, and $i-z$) diagrams of the cirrus clouds can be reproduced by scattered light if the interstellar radiation field (ISRF) of Mathis et al. (as updated by Draine) is modified, either by reducing the intensities in the $i$ and $z$ bands or by enhancing those in the $g$ and $r$ bands.
Similar results can also be obtained by adjusting the scattering albedos at the corresponding wavelengths.
This demonstrates that the color--color diagrams are effective not only for identifying extragalactic LSB features but also for studying the ISRF and the properties of interstellar dust.
\end{abstract}


\section{Introduction} 
Modern deep optical surveys have revealed faint tidal features around galaxies, enabling us to address key questions about their hierarchical evolution, which is intrinsically linked to the low surface brightness (LSB) Universe.
However, despite significant observational and technical advancements aimed at improving the quality of deep optical datasets, the widespread presence of interstellar material in our own Galaxy that reflects starlight continues to pose a major challenge \citep{Roman2020,Zhang2023,Liu2025}.
These filamentary structures of dust are called ``Galactic cirrus clouds.''
They were discovered in the far-infrared (FIR) by Infrared Astronomical Satellite (IRAS) observations in the early 1980s \citep{Low1984,Paley1991}.
These filamentary and clumpy clouds are also observable in both optical \citep{Sandage1976,deVries1985,Laureijs1987,Guhathakurta1989,Miville-Deschenes2016} and ultraviolet wavelengths \citep{Lee2008,Seon2011,Boissier2015} through the reflection of starlight.
A catalogue of dust clouds in the Galaxy, containing 525 high-latitude clouds, is given in \citet{Dutra2002}.
Because Galactic cirrus clouds can closely resemble the morphology and brightness of faint extragalactic features, they have become a major obstacle in modern optical surveys.

This problem is especially pronounced with today's deeper optical observations and persists even at high Galactic latitudes.
In particular, the Rubin Observatory, the Euclid mission, and the new generation of extremely large telescopes may encounter the ubiquitous presence of faint Galactic cirrus clouds covering the entire sky \citep{Ivezic2019_LSST,Euclid2025}.
Novel observing systems designed to achieve deep observations with a wide field of view have been developed.
The Dragonfly Telephoto Array is a unique system that utilizes multiple Canon 400 mm $f$/2.8 lenses \citep{Abraham2014}.
The Korea Astronomy and Space Science Institute (KASI) has been developing the KASI Deep Rolling Imaging Fast Telescope (K-DRIFT), a telescope optimized for LSB observations \citep{Byun2022, Byun2025, Ko2025}.
K-DRIFT is a wide-field optical telescope that employs off-axis freeform three-mirror optics to explore the southern hemisphere sky in the wavelength range 3000–7000\AA.
In observation with these systems, distinguishing extragalactic LSB objects from Galactic cirrus clouds is becoming increasingly essential.

FIR or submillimeter full-sky maps, such as those from the IRAS or the Planck missions \citep{Neugebauer1984_IRAS, PlanckCollaboration2014_I}, may be used to identify the Galactic cirrus clouds in the optical images of LSB objects.
However, the poor spatial resolution of FIR and submm images makes them inefficient for removing dust foreground contamination at smaller angular scales.
To overcome this difficulty, \citet{Roman2020} investigated the optical colors derived from $g$, $r$, $i$, and $z$ band observations in the Sloan Digital Sky Survey (SDSS) Stripe 82 region and found that the colors of Galactic cirrus clouds differ significantly from those of extragalactic sources.
For a given $g-r$ color, cirrus clouds exhibit a bluer $r-i$ color, allowing them to be distinguished from extragalactic objects.
They also provide an empirical criterion for identifying extragalactic objects: $(r-i) > 0.43\times(g-r) - 0.06$.
\citet{Mattila2023} also studied the colors of cirrus, translucent, and opaque dust clouds in the high-latitude area of LDN 1642.
They found that the color is bluer in low-column-density regions and redder in high-column-density regions.
This color variation with dust column density is a direct consequence of the wavelength dependence of extinction.
These observational results can be qualitatively understood by considering that dust-scattered light produces bluer colors, since starlight is more effectively scattered at shorter wavelengths \citep{Witt1992,Witt2000}.

In this study, we employ dust radiative transfer (RT) models to quantitatively investigate the colors of starlight scattered by optically thin Galactic cirrus clouds, with particular attention to those observed in the Stripe 82 region.
We find that the colors of dust-scattered light can serve as constraints on the spectral shape of the interstellar radiation field (ISRF) in the vicinity of the Sun.
Section \ref{sec:method} describes the assumptions of our simulations, the density field distribution, and the Monte Carlo simulation methods.
The main results are presented in Section \ref{sec:results}, while Section \ref{sec:discussion} addresses issues closely related to these results.
Finally, Section \ref{sec:summary} summarizes our findings and provides concluding remarks.

\section{Methods} \label{sec:method}

\subsection{Dust Medium}

The interstellar medium (ISM) is turbulent, resulting in a clumpy density distribution.
The statistical properties of a turbulent medium can be characterized by a power spectral density (PSD) in the Fourier domain and a probability density function (PDF) in real space.
The PSDs of turbulent media are well represented by power-law functions, and the PDFs of their three-dimensional (3D) densities are approximately lognormal, as demonstrated by both numerical simulations and observations \citep{Vazquez-Semadeni1994,Padoan2004,KimRyu2005,Berkhuijsen2008}.
The relationship between the power-law index of the PSD and the sonic Mach number ($M_{\rm s}$) of an isothermal turbulent medium was summarized in \citet{Seon2012} based on the numerical simulation results.

The ISM density fields can be effectively modeled using fractional Brownian motion, as detailed in \citet{Seon2012} and \citet{Seon2016}.
\citet{Seon2012} derived a relationship between the power-law index of a Gaussian random field and that of a lognormal random field generated from it, which is then used to calculate a lognormal random field with a desired power-law PSD.
\citet{Lewis2002} had developed an approximate iterative method for generating a lognormal density field with a prescribed power-law PSD.
We compared their method with ours and found that the two approaches produce statistically consistent results.
In our method, we adopted a cubic box of 128$^3$ uniform Cartesian cells to derive the formulae and confirmed the validity of the method up to a box size of 512$^3$.
However, for larger box sizes (256$^3$ or higher), we found that the method proposed by \citet{Seon2012} tends to produce slightly lower power at small spatial scales, although the results remain statistically valid.

We note that higher values of $M_{\rm s}$ correspond to clumpier media with greater density contrast; the variance ($\varsigma^2$) of the log-density ($\ln\rho$) is related to $M_{\rm s}$ by
\begin{equation}
\varsigma^2 = \ln(1+b^2 M_{\rm s}^2), \label{eq01}
\end{equation}
where the proportional constant $b$ is $\sim$0.4 for natural turbulence forcing mode \citep{Federrath2008}.
To examine the variability of the results across different realizations, 10 random realizations were generated for each $M_{\rm s}$.
Although slight variations appeared across realizations at large $M_{\rm s}$, no statistically significant variation were found.

In this study, we used 10 random realizations of lognormal density fields with a box size of $128^3$, generated using the method of \citet{Seon2012}, for each Mach number ($M_{\rm s}=$ 0.5, 1, 2, 4, 6, 8, 10, 12, 16, and 20).
Examples of such density fields are shown in \citet{Seon2016}.

\begin{figure}[t!]
\begin{center}
\includegraphics[width=0.46\textwidth]{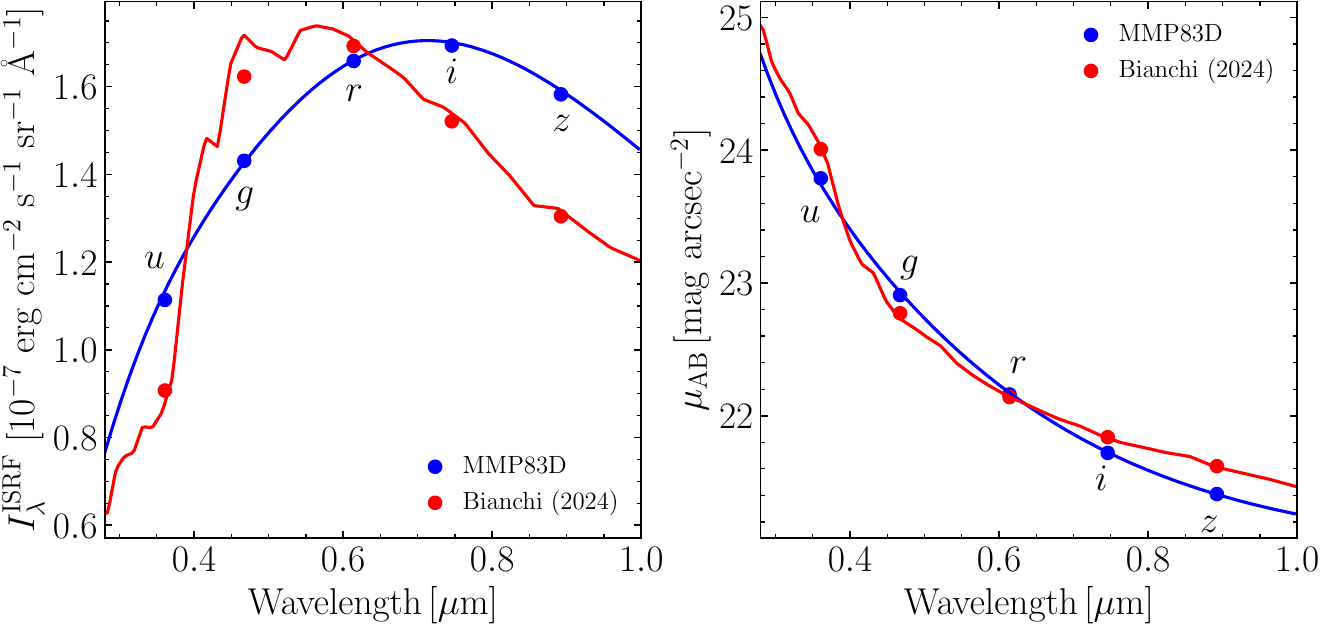}
\end{center}
\caption{SEDs of the two ISRF models. The left and right panels show the SEDs in surface brightness units of erg cm$^{-2}$ s$^{-1}$ sr$^{-1}$ \AA$^{-1}$ and mag arcsec$^{-2}$, respectively.
The models of \citet{Mathis1983} (as updated by \citealt{Draine2011book}) and \citet{Bianchi2024} are plotted as blue and red curves, respectively.
The blue and red filled circles indicate the intensities at the central wavelengths of the SDSS bands, calculated by weighting with the SDSS filter functions.
\label{fig01}}
\end{figure}

\subsection{Interstellar Radiation Field}

For the spectrum of the ISRF incident on the cloud, we adopted two models: (1) the spectral energy distribution (SED) originally proposed by \citet{Mathis1983} and later modified by \citet{Draine2011book} to better match the COBE-DIRBE photometry; and (2) the SED recently derived by \citet{Bianchi2024}.
In this paper, we refer to the ISRF of \citet{Mathis1983}, updated by \citet{Draine2011book}, as MMP83D.
The SEDs at optical wavelengths are shown in Figure \ref{fig01}.
In the left panel, the SEDs ($I^{\rm ISRF}_\lambda$) are plotted in intensity units of erg cm$^{-2}$ s$^{-1}$ sr$^{-1}$ $\rm\AA^{-1}$, while in the right panel they are shown in units of mag arcsec$^{-2}$.
The figure also indicates the central wavelengths of the five SDSS bands---$u$, $g$, $r$, $i$, and $z$---along with the mean intensities in each band, calculated by weighting the ISRF spectrum with the corresponding filter functions.
As shown in the figure, the model of \citet{Bianchi2024} exhibits lower intensities in the $u$, $i$, and $z$ bands and higher intensity in the $g$ band, while in the $r$ band the two models approximately agree.

In this study, we calculated the intensities of scattered light based on the mean intensities in the SDSS bands, as illustrated in Figure~\ref{fig01}.
We also note that the total luminosity incident on the six surfaces of the box is given by
\begin{equation}
\mathcal{L}_\lambda=\pi I_\lambda^{\rm ISRF}\times 6 L^2, \label{eq02}
\end{equation}
where $L$ is the length of one side of the box, assuming an isotropic radiation field.
All results in this paper are normalized to the incident intensity ($I^{\rm ISRF}_\lambda$).

\begin{figure*}[ht!]
\begin{center}
\includegraphics[width=\textwidth]{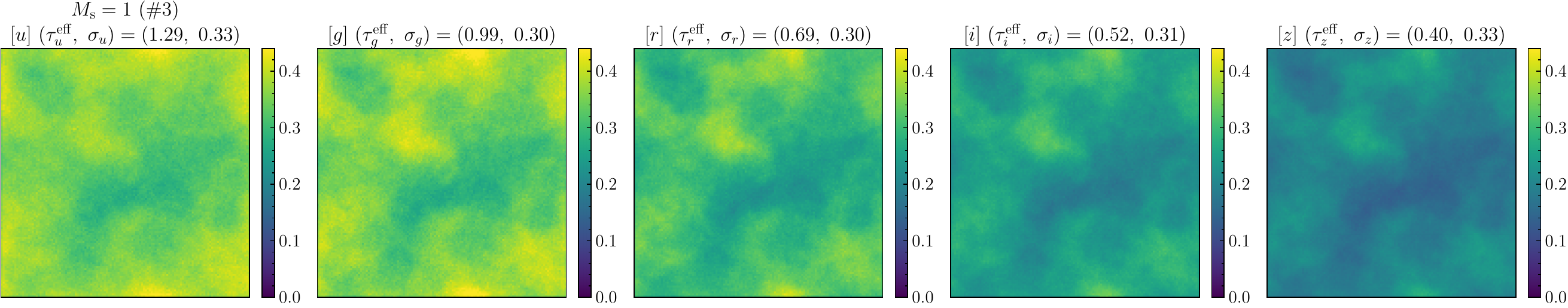}\vspace{0.1cm}
\includegraphics[width=\textwidth]{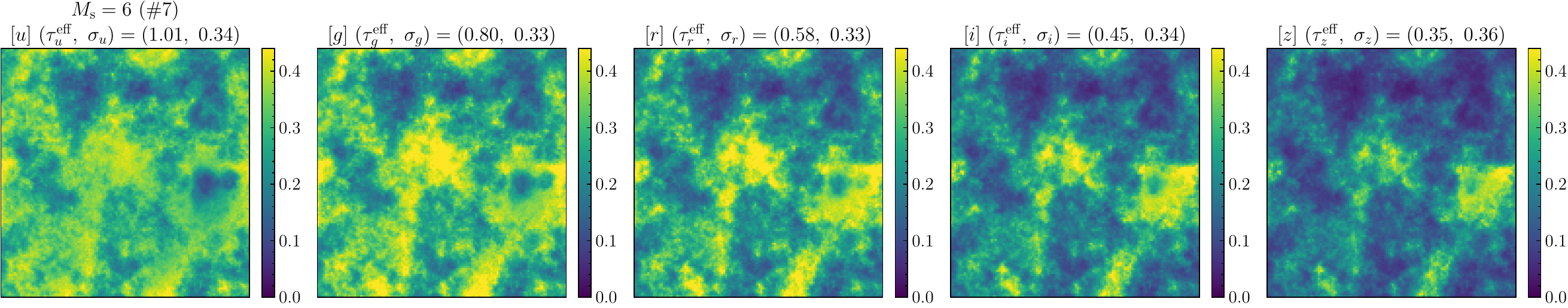}
\end{center}
\caption{Maps of scattered light in the SDSS wavelength bands ($u$, $g$, $r$, $i$, and $z$) for two realizations with different Mach numbers: $M_{\rm s}=1$ (top panels) and $M_{\rm s}=6$ (bottom panels). The homogeneous optical depth in the $g$ band is assumed to be $\tau_g^{\rm h}=1.0$ in both cases. The maps are based on two typical random realizations of a lognormal density structure. The numbers (\#3 and \#7) in parentheses indicate the sample indices among the 10 random realizations of the density distribution. Scattered light intensities are normalized by the incident ISRF intensity, i.e., $I_{\rm scatt}/I_{\rm ISRF}$. The effective optical depth $\tau^{\rm eff}_\lambda$, defined in Equation (\ref{eq04}), and the standard deviation $\sigma_\lambda$ of scattered light (for $\lambda = u$, $g$, $r$, $i$, and $z$) are indicated in each map.
\label{fig02}}
\end{figure*}

\subsection{Simulations}

Simulations were conducted using the Monte Carlo RT code MoCafe 
\citep{Seon2013,Seon2014,Seon2015,Seon2016}.
The code is parallelized with the Message Passing Interface.
The Henyey-Greenstein function was adopted to describe the dust-scattering phase function \citep{Henyey1941,Witt1977}.
The extinction cross section, albedo ($a$), and asymmetry factor ($g\equiv\left<\cos\theta\right>$), all of which are wavelength-dependent, for the SDSS bands were taken from the theoretical dust model of \citet{Weingartner2001} and \citet{Draine2003ApJ}.
The extinction cross sections relative to the $g$ band are given by 1.32, 1.00, 0.70, 0.53, and 0.40 for the $u$, $g$, $r$, $i$, and $z$ bands, respectively.
The albedos are 0.62, 0.66, 0.68, 0.67, and 0.64, respectively, and the corresponding asymmetry factors are 0.57, 0.56, 0.51, 0.46, and 0.40.
The first scattering of each photon is forced to improve the efficiency of RT simulations in optically thin media.
The code has previously been applied to RT calculations with internal radiation sources.
In this study, it was updated to handle cases in which the medium is externally illuminated by background ambient radiation.

The RT simulations were performed by scaling the density fields, parameterized by the homogeneous optical depth $\tau_g^{\rm h}$ (also referred to as the cloud optical depth).
This parameter is defined as the $g$-band optical depth, measured across a uniform medium of the same total mass, from one boundary wall to the opposite side.
The homogeneous optical depth $\tau_g^{\rm h}$ was varied from 0.2 to 4.0 in steps of 0.2 for each realization at each Mach number and for each SDSS band.
In total, 10,000 simulations were performed, using 10$^8$ photon packets for each simulation.

The radiation field is assumed to be isotropically incident on the surfaces of the cloud.
The location where a photon enters the cloud is chosen uniformly at random from one of the six boundary surfaces of the box.
The incident direction of the photon is determined by a polar angle ($\vartheta$), given by $\mu \equiv \cos\vartheta = -\xi^{1/2}$, and an azimuthal angle ($\varphi$), given by $\varphi = 2\pi\xi'$, both defined with respect to the surface normal (pointing outward from the cloud) at the point of incidence.
Here, $\xi$ and $\xi'$ are two independent pseudo-random numbers uniformly distributed between 0 and 1.
The polar angle is obtained from the condition of isotropic intensity, $\xi = 2\int_0^\mu \mu' d\mu'$.
Finally, the photon direction vector is transformed into the coordinate system of the cloud.

The peel-off technique (more precisely, next-event estimation) is applied to enable the calculation of two-dimensional images as they would be detected by an observer located at a position and orientation (see \citealt{Yusef1984}, \citealt{Noebauer2019}, and \citealt{Seon2022} for general descriptions of the technique).
At each scattering event, the peel-off component for scattered light is obtained by calculating the probability that the photon packet is scattered toward the observer.

In MoCafe, at each incident-photon event, a `virtual' photon packet is generated at the same boundary location as the incident photon, and its probability of reaching the observer defines the direct radiation component.
The probability is given by
\begin{equation}
p_{\rm direct}=\frac{\left|\cos\vartheta\right|}{\pi d^2}e^{-\tau}, \label{eq03}
\end{equation}
where $\vartheta$ ($\pi/2 \le \vartheta \le \pi$) is the angle between the direction of the virtual photon and the surface normal at the point of incidence (pointing outward from the cloud); $d$ is the distance from the photon to the observer; and $\tau$ is the optical depth along the line of sight through the cloud, measured from the photon's initial location on the cloud surface.
It is important to note that when $\cos\vartheta > 0$, the probability is zero, as no radiation is emitted from the cloud.
In the above equation, the $\pi$ in the denominator is the normalization factor of the probability, given by $ \int_{0}^{2\pi} \int_{\pi/2}^{\pi}\cos\vartheta \sin\vartheta d\vartheta d\varphi$. The term $1/d^2$ represents the geometrical dilution factor, and $e^{-\tau}$ denotes the fraction of photon packet that passes through the medium without being extinguished.
The peel-off components collected at a pixel on the detector plane are divided by the total number of photon packets and its solid angle subtended by the cloud, and then multiplied by the total ISRF luminosity $\mathcal{L}_\lambda$ incident on the cloud.
The code also calculates the intensity of the direct radiation expected in the absence of extinction by setting $\tau=0$ in Equation (\ref{eq03}).
This component was confirmed to be uniform across the detector plane and to match the intensity of the ISRF, $I_\lambda^{\rm ISRF}=\mathcal{L}_\lambda/(6 L^2\times \pi)$.

\subsection{Observational Data}
The present simulation results are compared with observations of cirrus clouds in the SDSS Stripe 82 region, measured in the $g$, $r$, $i$, and $z$ bands, as reported by \citet{Roman2020}.
In observational data analyses, differential photometry is applied, where the signal from the target is subtracted from the adjacent background sky signal \citep{Witt2008,Mattila2018}.
Consequently, the resulting measurement does not represent purely scattered light.
However, as discussed in Section \ref{sec:discussion}, this effect is likely of minor importance, particularly for the case considered in this study.

\begin{figure*}[ht!]
\begin{center}
\includegraphics[width=\textwidth]{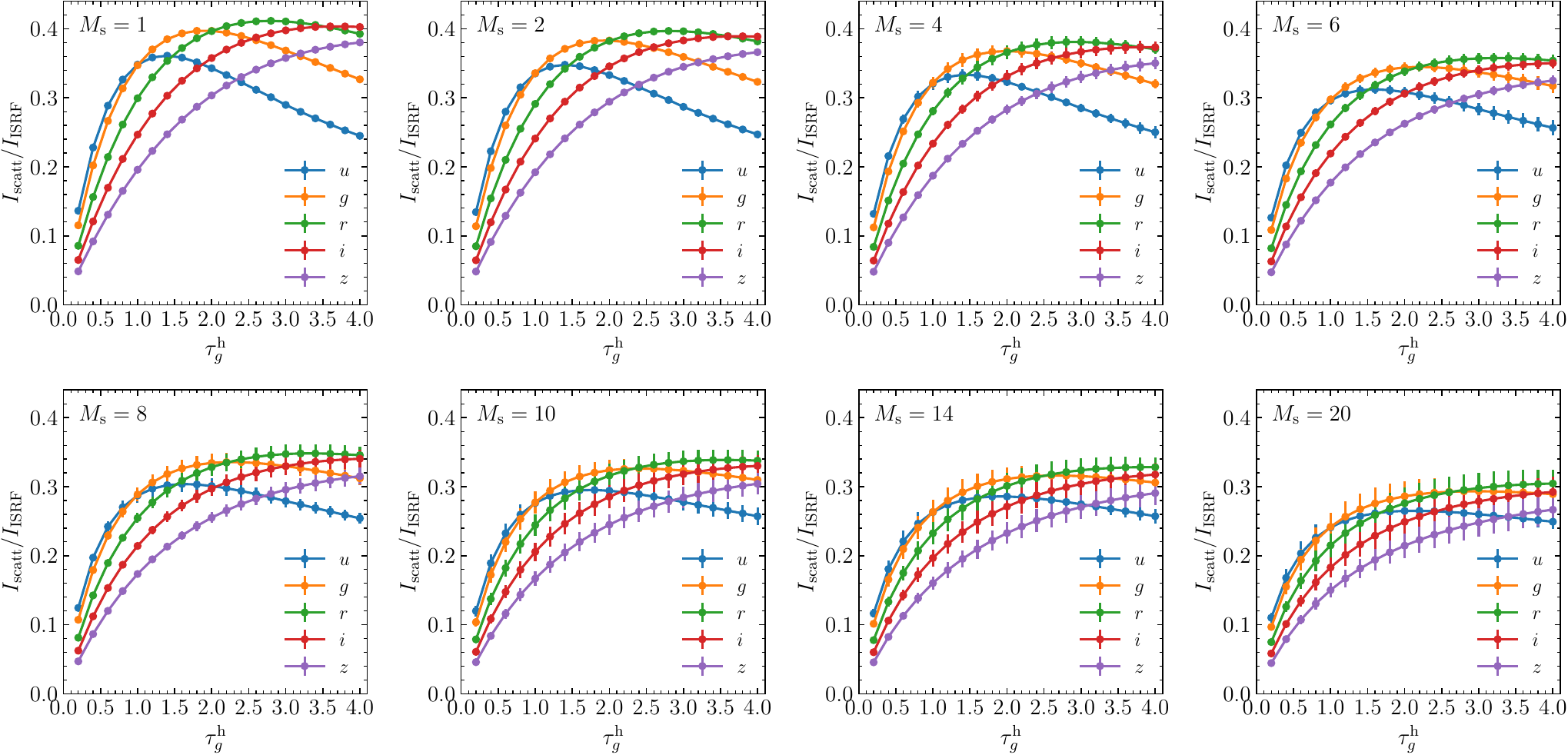}
\end{center}
\caption{Mean scattered-light intensities as functions of $\tau_g^{\rm h}$ for various $M_{\rm s}$. The intensities are averaged over 10 random realizations of the density structure in the RT simulations. Standard deviations across the 10 samples are shown as error bars.
\label{fig03}}
\end{figure*}

\section{Results} \label{sec:results}

\subsection{Overall Properties}

We first present example maps of scattered light. Figure \ref{fig02} shows such maps for two Mach numbers, $M_{\rm s}=1$ and $M_{\rm s}=6$, across five wavelength bands.
The results were obtained assuming a homogeneous optical depth in the $g$ band of $\tau_{\rm g}^{\rm h}=1.0$.
For reference, the effective optical depth, defined as
\begin{equation}
\tau^{\rm eff}_\lambda = -\ln\left<e^{-\tau_\lambda^{\rm los}}\right>, \label{eq04}
\end{equation}
is also shown in each panel.
Here, $\tau_\lambda^{\rm los}$ represents the optical depth along each line of sight (i.e., each pixel) in the map, and the angle brackets denote the average over all pixels in the image.
The effective optical depths are consistently smaller than the homogeneous optical depth at the corresponding wavelengths \citep{Witt1996,Seon2016}.
The departure increases with Mach number, i.e., as the medium becomes clumpier.
This results from the statistical properties of the lognormal distribution.
For a fixed total mass of the medium (i.e., a fixed $\tau_g^{\rm h}$), an increase in density dispersion (i.e., $M_{\rm s}$) reduces the volume fraction of high-density regions while increasing the maximum density.
Simultaneously, the volume fraction of low-density regions increases.
As a result, media with larger $M_{\rm s}$ values exhibit lower effective optical depths.

The maps for the higher Mach number ($M_{\rm s}=6$, bottom panels) are noticeably clumpier and more complex than those for the lower Mach number ($M_{\rm s}=1$, top panels), reflecting the inherently more structured and fragmented nature of high-Mach-number turbulence.
In general, scattered light is brighter at shorter wavelengths, except in the $u$-band image, which appears slightly fainter than the $g$-band image.
These properties are more clearly illustrated in Figure~\ref{fig03}, which shows the intensity as a function of optical depth in each band.
It is also evident that the maps at different wavelengths are generally similar, but they exhibit differences in detail.
The standard deviation $\sigma_\lambda$, which quantifies the contrast, is shown in each map.
The scattered light generally exhibits greater contrast at longer wavelengths, particularly at higher $M_{\rm s}$.
This trend, evident in all bands except $u$, results from the increased optical depth at shorter wavelengths: photons are scattered even in the lowest-density regions, and repeated scatterings smooth the resulting maps.
At longer wavelengths, by contrast, scattering is less frequent and occurs mainly in the highest-density regions, producing maps with higher contrast.
In the optically thick $u$ band, however, most of the observed scattered light arises from single scattering near the outer boundaries, yielding high contrast in opposition to the trend found in other bands.
Furthermore, in some regions, the longer-wavelength intensity is anti-correlated with the shorter-wavelength intensity, particularly at high Mach numbers (i.e., $M_{\rm s}=6$).
This anti-correlation occurs along sightlines that are optically thick at shorter wavelengths but optically thin at longer wavelengths, when the line-of-sight dust column density lies within an intermediate range.

\begin{figure*}[ht!]
\begin{center}
\includegraphics[width=\textwidth]{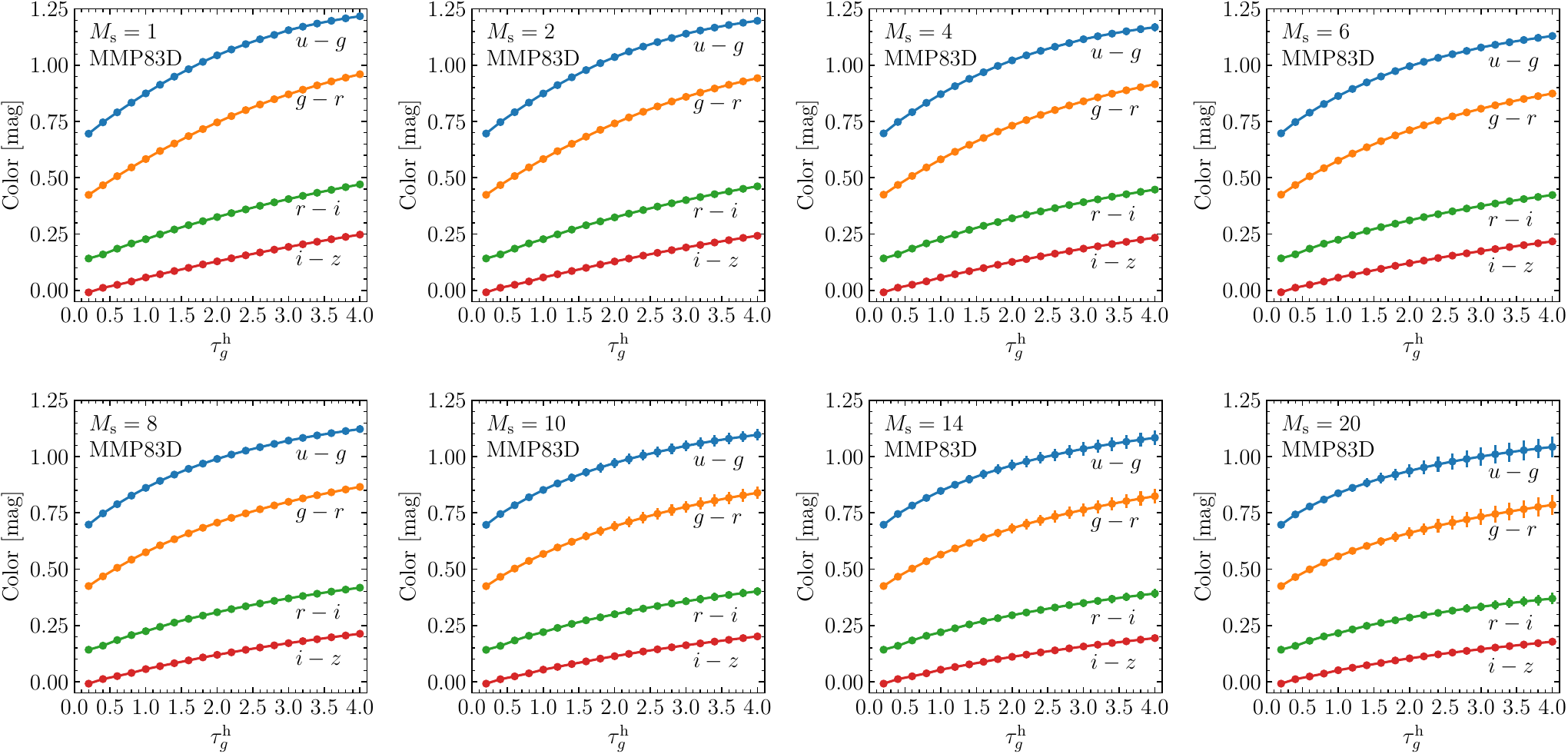}
\end{center}
\caption{Mean colors ($u-g$, $g-r$, $r-i$, and $i-z$) as functions of $\tau_g^{\rm h}$ for various $M_{\rm s}$. The values are averaged over 10 random realizations of the density structure in the RT simulations. Standard deviations are also shown as error bars. The ISRF SED of MMP83D was used.
\label{fig04}}
\end{figure*}

In Figure \ref{fig03}, we present the scattered intensity, averaged over the entire image regions in five bandpasses, as a function of the homogeneous optical depth ($\tau_g^{\rm h}$) for various Mach numbers ($M_{\rm s}$).
The scattered intensity is expressed relative to the incident ISRF intensity and is calculated as the mean over 10 random realizations for each Mach number.
The standard deviations, derived from the 10 realizations, are also shown as error bars.
It is notable that the scattered light reaches up to approximately 40\% of the incident light, particularly in the case of $M_{\rm s}=1$.
As shown in Figure \ref{fig02}, the normalized scattered intensity ($I_{\rm scatt}/I_{\rm ISRF}$) increases as wavelength decreases in the optically thin regime ($\tau_g^{\rm h} \lesssim 1$).
However, in the optically thick regime, scattered light decreases rapidly in the short-wavelength bands.
In general, both the maximum intensity and the overall level of scattered light decrease with increasing $M_{\rm s}$.
To understand this behavior, it is important to note that scattering occurs in proportion to the optical depth in the optically thin regime but becomes suppressed in the optically thick regime \citep{Witt2010,Mattila2018,Mattila2023}.
This tendency is evident in the $u$ and $g$ bands at relatively low Mach numbers; however, it is less pronounced in the other bands, as they generally remain in the optically thin regime at longer wavelengths.
Additionally, as described above, in a clumpier medium with a higher $M_{\rm s}$, the volume fraction of low-density regions increases, whereas high-density regions occupy a smaller volume.
As a result, scattering tends to occur more prominent in less clumpy media---that is, media with lower Mach numbers.
The dispersion in scattered light across different density realizations also increases at higher Mach numbers due to greater voxel-to-voxel density variation.

\begin{figure*}[ht!]
\plotone{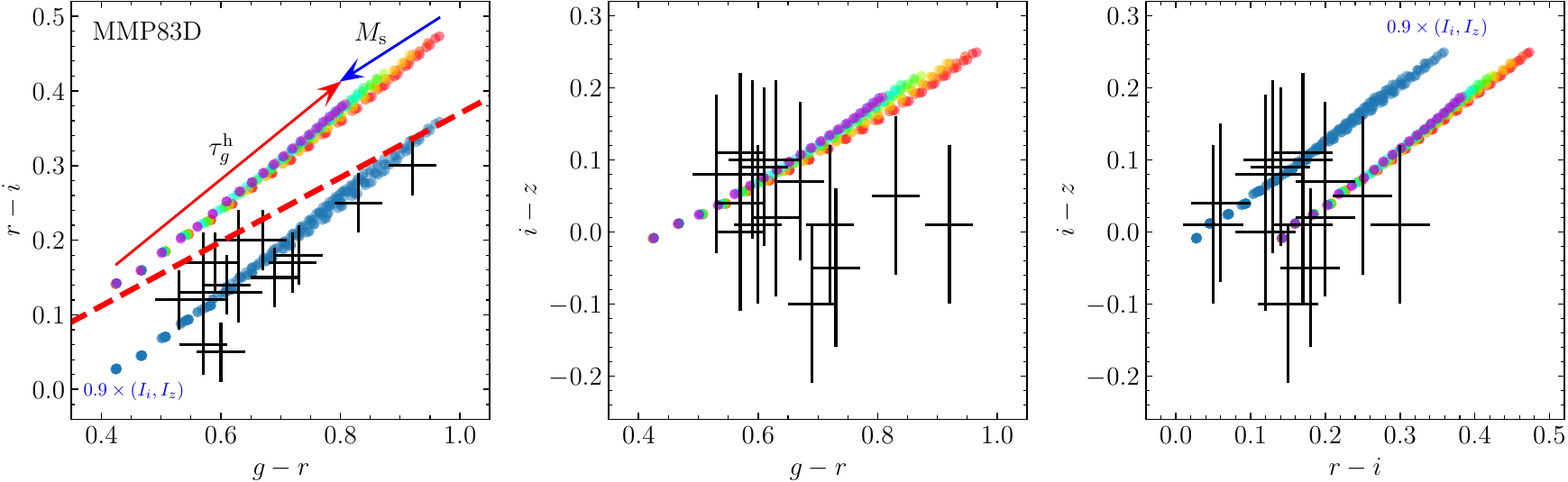}
\caption{Mean color--color diagrams constructed for $\tau_g^{\rm h}=0.2$--4.0 and $M_{\rm s} = 0.5$--20. Rainbow-colored dots show the simulation results with the unmodified MMP83D ISRF, while light-blue dots represent cases where the incident ISRF in the $i$ and $z$ bands is scaled down by a factor of 0.9 to reproduce the observed data. Each dot represents the average across 10 realizations for a given combination of $\tau_g^{\rm h}$ and $M_{\rm s}$, as in Figures \ref{fig03} and \ref{fig04}. In the rainbow-colored series, each color corresponds to a fixed $M_{\rm s}$ as $\tau_g^{\rm h}$ varies. The red arrow indicates the color trend with increasing $\tau_g^{\rm h}$, while the blue arrow show the trend with increasing $M_{\rm s}$. The black crosses mark the observational results of \citet{Roman2020}. In the left panel ($r-i$ vs.\ $g-r$), the red dashed line denotes the empirical relation used by \citet{Roman2020} to separate cirrus clouds from extragalactic LSB features. The reddening with increasing $\tau_g^{\rm h}$ is due to stronger extinction at shorter wavelengths, whereas the blueing with increasing $M_{\rm s}$ results from the larger volume fraction of low-density regions.
\label{fig05}}
\end{figure*}

Figure \ref{fig04} shows the variations in the colors $u-g$, $g-r$, $r-i$, and $i-z$ with increasing optical depth $\tau_g^{\rm h}$.
The ISRF SED of MMP83D was used for this calculation.
In general, colors involving shorter wavelengths are larger than those involving longer wavelengths, because short-wavelength photons are more strongly attenuated by dust through scattering and absorption.
The colors correlate approximately linearly with optical depth in optically thin regimes ($\tau_g^{\rm h}\lesssim2$), but the correlation weakens with increasing $\tau_g^{\rm h}$ and $M_{\rm s}$.
The slope is steeper at shorter wavelengths and shallower at longer wavelengths.
This correlation indicates reddening with increasing depth, and steeper slopes at shorter wavelengths indicate faster reddening.
We also note that, for a fixed $\tau_g^{\rm h}$, the colors decrease and thus become bluer as $M_{\rm s}$ increases.
This blueing is attributable to increased density dispersion at higher $M_{\rm s}$, which enhances the scattering of shorter wavelengths in low-density regions that progressively occupy larger volumes.

These results agree with the observations of \citet{Roman2020}, who reported that the correlation between cirrus optical colors and the dust column density (defined as the optical depth at 353 GHz, $\tau_{353}$) is stronger for colors involving shorter wavelengths.
Unlike their study, which found almost no correlation between $i-z$ and $\tau_{353}$, our results reveal a weak correlation.
However, given the large uncertainties in the observed $i-z$ colors, the difference may not be statistically significant.

\begin{figure*}[ht!]
\plotone{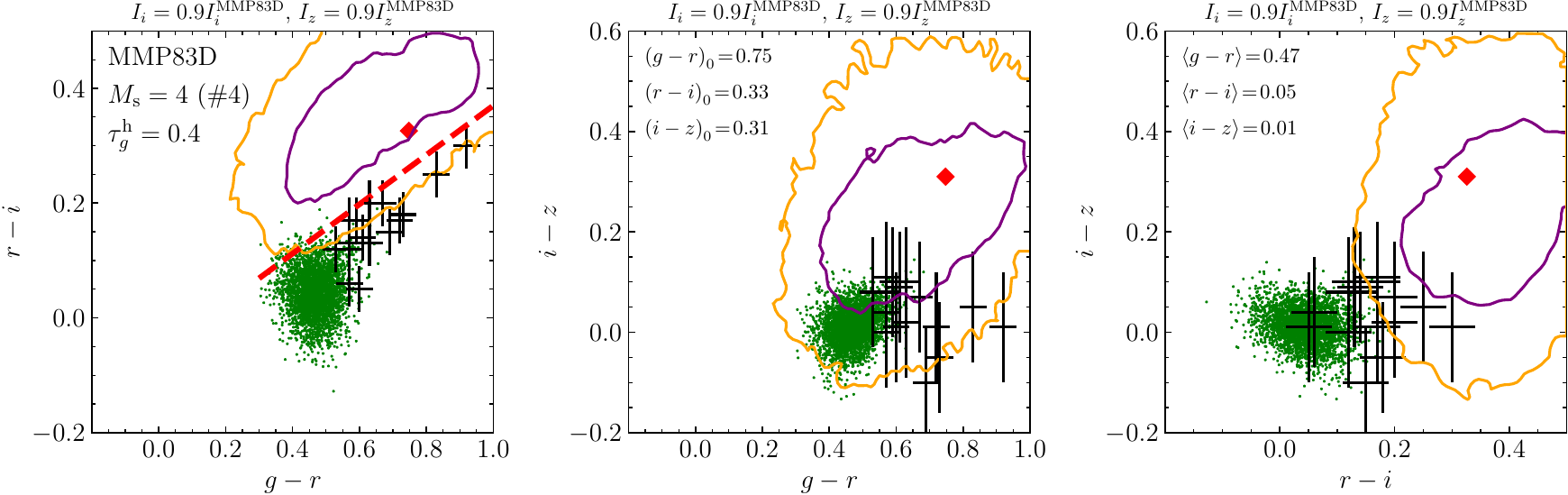}\vspace{0.3cm}
\plotone{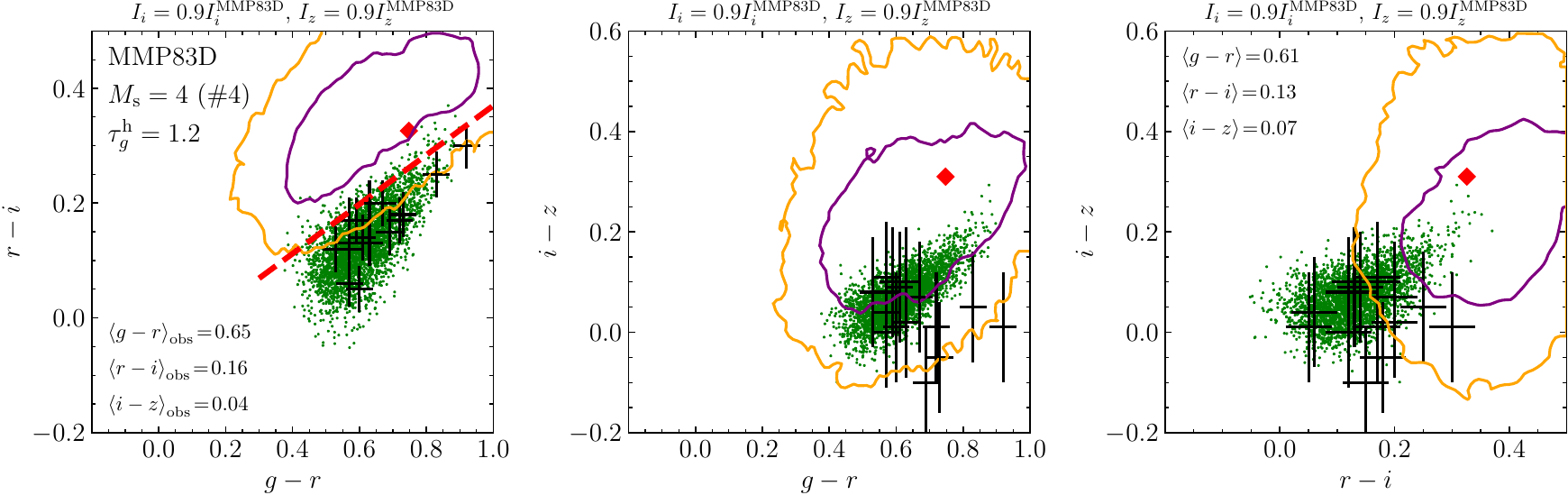}\vspace{0.3cm}
\plotone{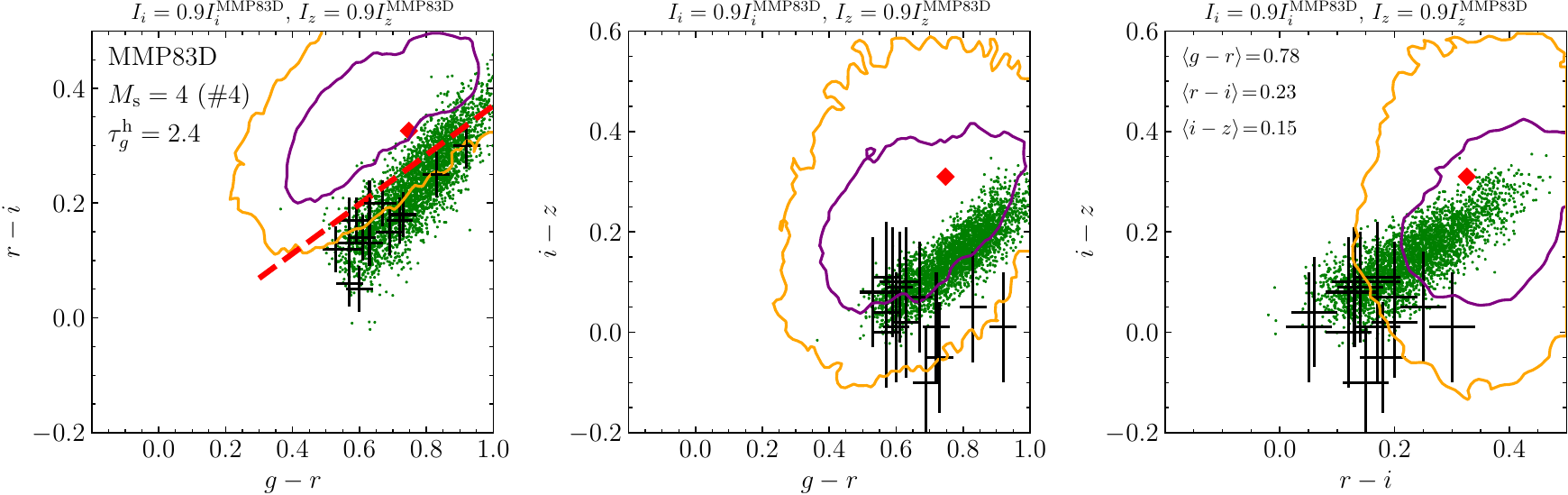}
\caption{color--color diagram variations as $\tau_g^{\rm h}$ changes. The optical depths are set to 0.4, 1.2, and 2.4 in the top, middle, and bottom panels, respectively, with the Mach number fixed at $M_{\rm s} = 4$.  The ISRF model of MMP83D is assumed, with the $i$- and $z$-band intensities scaled down by a factor of 0.9 to match the $r-i$ color. Green dots show the simulation results of scattered light, and red diamonds indicate the colors of the incident ISRF, and black crosses mark the observational results of \citet{Roman2020}. Yellow and purple contours represent regions enclosing the 1$\sigma$ and 2$\sigma$ ranges of external galaxy samples from \citet{Roman2020}. The red dashed lines in the left panels ($r-i$ vs.\ $g-r$) denote the empirical relation used to distinguish cirrus clouds from extragalactic LSB features \citep{Roman2020}. The initial colors of the input ISRF (MMP83D) are labeled with a subscript `0' in the top-middle panel. Median values of the three simulated colors are shown in the right panels, while the mean observed colors are shown with a subscript `obs' in the middle-left panel. Note that the medians of the observed colors equal the means, except for the $g-r$ color, whose median (0.62) is slightly smaller. The number (\#4) in parentheses indicates the random sample index among the 10 realizations.
\label{fig06}}
\end{figure*}

\begin{figure*}[ht!]
\plotone{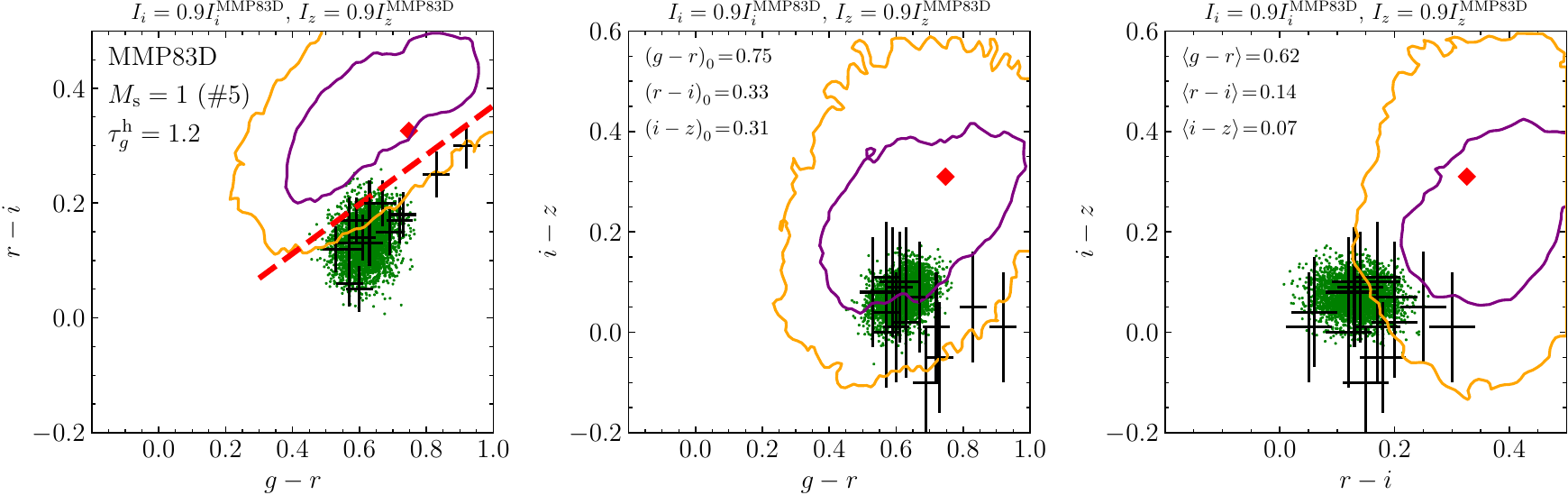}\vspace{0.3cm}
\plotone{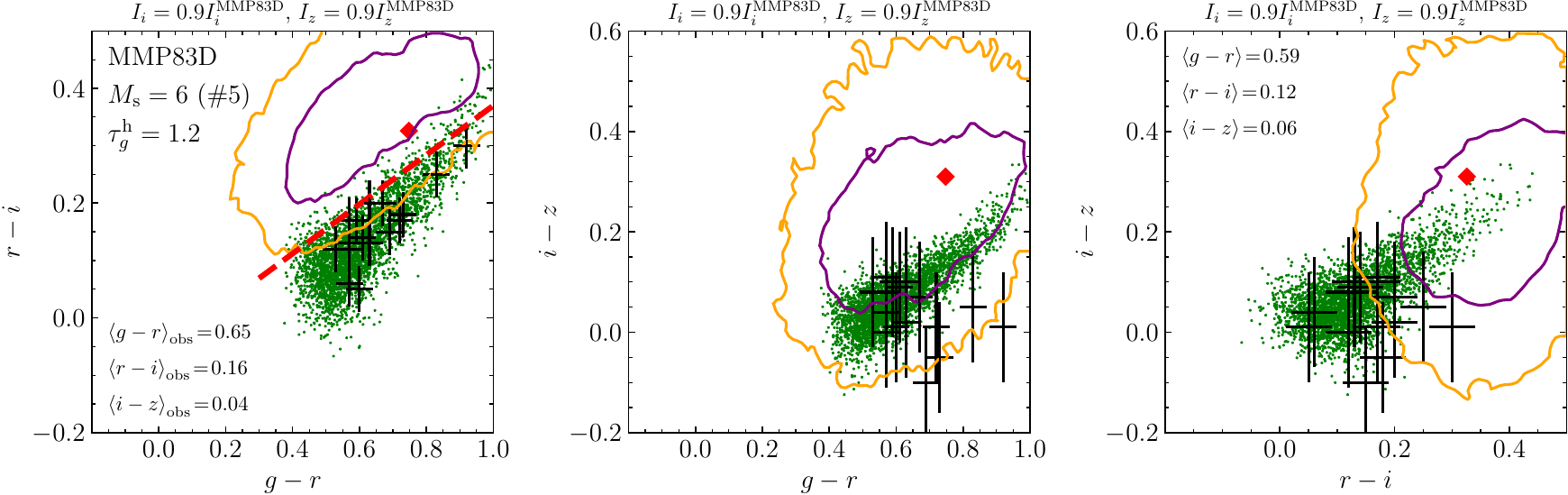}\vspace{0.3cm}
\plotone{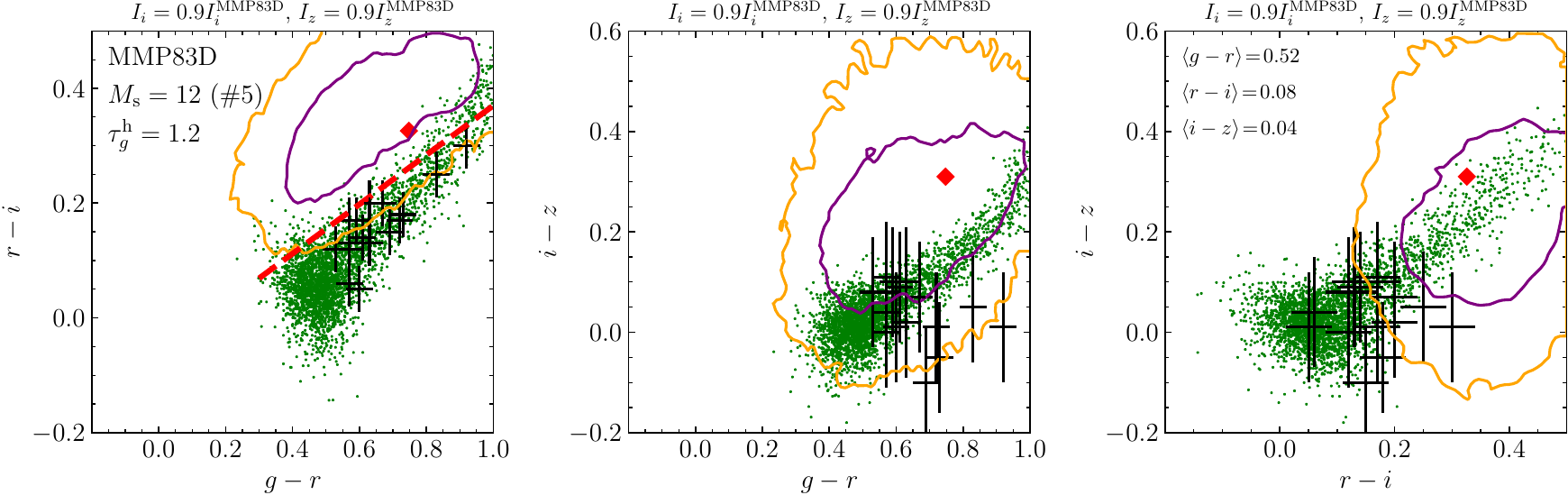}
\caption{color--color diagram variations with changing $M_{\rm s}$. The Mach numbers are set to 1, 6, and 12 in the top, middle, and bottom panels, respectively, with the cloud optical depth fixed at $\tau_g^{\rm h} = 1.2$. The modified SED of the MMP83D SED was assumed. Green dots show the simulation results of scattered light, while red diamonds indicate the colors of the incident ISRF. Observational results from the Stripe 82 region and external galaxy samples from \citet{Roman2020} are also shown, as in Figure \ref{fig06}. The initial ISRF colors, the median simulated colors, and the mean observed colors are indicated.
\label{fig07}}
\end{figure*}

\subsection{Color--Color Diagrams}

In this section, we first examine the overall properties of the optical color--color diagrams (defined by $g-r$, $g-i$, and $i-z$) obtained from simulations and then compare them with observations of the Stripe 82 region by \citet{Roman2020}.

Figure \ref{fig05} shows the color--color diagrams based on the mean colors, which were first averaged over individual scattered-light maps and then further averaged across 10 realizations for each $\tau_g^{\rm h}$ and $M_{\rm s}$.
The ISRF model of MMP83D was adopted as in Figure \ref{fig04}.
In the rainbow-colored dots, $M_{\rm s}$ increases with the color transition from red to purple.
The red arrow illustrates the track formed as $\tau_g^{\rm h}$ increases from 0.2 to 4.0 at a fixed $M_{\rm s}$ (with the highest value, $M_{\rm s}=20$), while the blue arrow marks the positions of the highest $\tau_g^{\rm h}$ (= 4.0) as $M_{\rm s}$ increases from 0.5 to 20.
The mean colors redden with increasing $\tau_g^{\rm h}$, forming a track for each $M_{\rm s}$, as indicated by the red arrow. Tracks corresponding to different $M_{\rm s}$ values largely overlap in the lower range of $\tau_g^{\rm h}$.
As $M_{\rm s}$ increases, however, the endpoint associated with the highest $\tau_g^{\rm h}$ shifts toward bluer colors (downward and to the left in the figure), leading to a shortened track.
In other words, while the starting point with the lowest $\tau_g^{\rm h}$ remains nearly fixed, the endpoint with the highest $\tau_g^{\rm h}$ is displaced along the track of a lower $M_{\rm s}$ toward smaller $\tau_g^{\rm h}$.

To understand the reddening with increasing $\tau_g^{\rm h}$, note that the rate of increase in scattered light with optical depth slows at relatively high optical depths, with the slowdown occurring more rapidly at shorter wavelengths, as shown in Figure \ref{fig03}.
At even higher optical depths, the scattered light begins to decrease.
This behavior results in the relative reddening of scattered light as optical depth increases, since shorter wavelengths experience higher extinction than longer wavelengths.
The blueing with increasing $M_{\rm s}$ arises from the growing density dispersion, which enhances the volume fraction of lower-density regions that contribute bluer colors to the scattered light.
The same effects were also described in relation to Figure \ref{fig04}.

In Figure \ref{fig05} and subsequent figures, the black crosses represent the observational data from \citet{Roman2020}.
In the left panel, the red dashed lines indicate the division between cirrus clouds and extragalactic features, as suggested by \citet{Roman2020}:
\begin{equation}
(r-i)=0.43\times(g-r)-0.06. \label{eq05}
\end{equation}
For a given $g-r$ color, the $r-i$ colors of cirrus clouds lie below those of extragalactic features.
In the figure, the predicted color--color diagrams of $g-r$ versus $i-z$ (second panel) and $r-i$ versus $i-z$ (third panel) encompass the observed results.
However, in the first panel, the predicted $r-i$ colors are larger than the observed values by $\sim$$0.1$ dex for a given $g-r$ color.

This mismatch in the $g-r$ versus $r-i$ diagram can be reduced by decreasing the $r-i$ color.
Such an adjustment can be achieved either by decreasing the $i$-band magnitude or by increasing the $r$-band magnitude.
At the same time, to keep the $i-z$ and $g-r$ colors unchanged, the $z$-band (or $g$-band) magnitude should be decreased by the same amount as the $i$-band (or $r$-band) magnitude.
Figure \ref{fig05} shows that when the ISRF intensities in the $i$ and $z$ bands are reduced by a factor of 0.9, the RT results (light blue dots) agree well with the observed color--color diagrams.
By definition, these color--color diagrams are identical to those obtained by scaling up the $g$- and $r$-band intensities by the inverse of this factor.
Therefore, we conclude that the observed color--color diagrams are reasonably well reproduced by modifying the SED of the MMP83D ISRF in the $i$ and $z$ bands (or equivalently in the $g$ and $r$ bands).
An alternative explanation would be to adjust the albedos by the same factor applied to the ISRF SED, either decreasing them in the $i$ and $z$ bands or increasing them in the $g$ and $r$ bands, since the scattered light intensity is roughly proportional to the albedo in optically thin cases.

Now, we examine the color--color diagrams obtained for individual density media as the optical depth ($\tau_g^{\rm h}$) and Mach number ($M_{\rm s}$) vary, and compare the results with the observations of \citet{Roman2020}.
Figure \ref{fig06} shows how the color--color diagrams change as $\tau_g^{\rm h}$ increases from 0.4 to 2.4, with the Mach number fixed at $M_{\rm s}=4$.
In this figure, $M_{\rm s}=4$ was chosen because it reproduces the observed dispersion in colors reasonably well.
The green dots represent the results when the MMP83D ISRF is assumed, after scaling the intensities in the $i$ and $z$ bands by a factor of 0.9.
The black crosses and red dashed lines indicate the observed data and the division line for extragalactic features, respectively, as in Figure \ref{fig05}.
The color--color distributions of external galaxies given by \citet{Roman2020} are also shown as yellow and purple contours.
In the right panels, the median values of the three colors ($\left<g-r\right>$, $\left<r-i\right>$, and $\left<i-z\right>$) calculated from simulations are presented.
The medians represent the typical colors better than the mean values because the color distributions are not symmetric.

As seen in the figure, the median colors become redder as the optical depth of the medium increases, and the mean colors show the same trend.
This behavior can be understood in the same way as described above for Figure \ref{fig05}.
At the same time, the figure also shows that the dispersion in color increases with $\tau_g^{\rm h}$, primarily due to the intrinsic properties of the lognormal distribution.
In a lognormal density distribution, the standard deviation $\sigma$ of the density $\rho$ is given by
\begin{equation}
\sigma = [\exp(\varsigma^2)-1]^{1/2}\mu, \label{eq06}
\end{equation}
where $\varsigma$ is the standard deviation of $\ln\rho$ and $\mu$ is the mean of $\rho$.
In other words, the dispersion of the density increases with the mean density.
Consequently, an increase in the cloud optical depth $\tau_g^{\rm h}$ (and hence in $\mu$) results in greater dispersion in the line-of-sight optical depth, and, in turn, in the colors.

Figure \ref{fig07} presents the changes in the color--color diagrams with increasing $M_{\rm s}$, while $\tau_g$ is fixed at 1.2.
The median colors are also shown in the right panels.
Interestingly, in this case, the median (and mean) colors shift slightly toward the blue with increasing $M_{\rm s}$---unlike in Figure \ref{fig06}, where increasing $\tau_g^{\rm h}$ leads to a redward shift---while the color dispersion increases.
This blueing effect is due to an increase in clumpiness, as described in connection with Figure \ref{fig05}.

The increase in color dispersion with rising $\tau_g^{\rm h}$ (Figure \ref{fig06}) arises from a relatively uniform spread of the colors, whereas the increase with $M_{\rm s}$ (Figure \ref{fig07}) is dominated by an extended redward tail in the color distribution.
This reflects the fact that variations in $\tau_g^{\rm h}$ simply scale the density, while changes in $M_{\rm s}$ modify its clumpiness.
Similar trends to those shown in Figures \ref{fig06} and \ref{fig07} are found for other combinations of $\tau_g^{\rm h}$ and $M_{\rm s}$, though they are not shown here.

\begin{figure*}[ht!]
\plotone{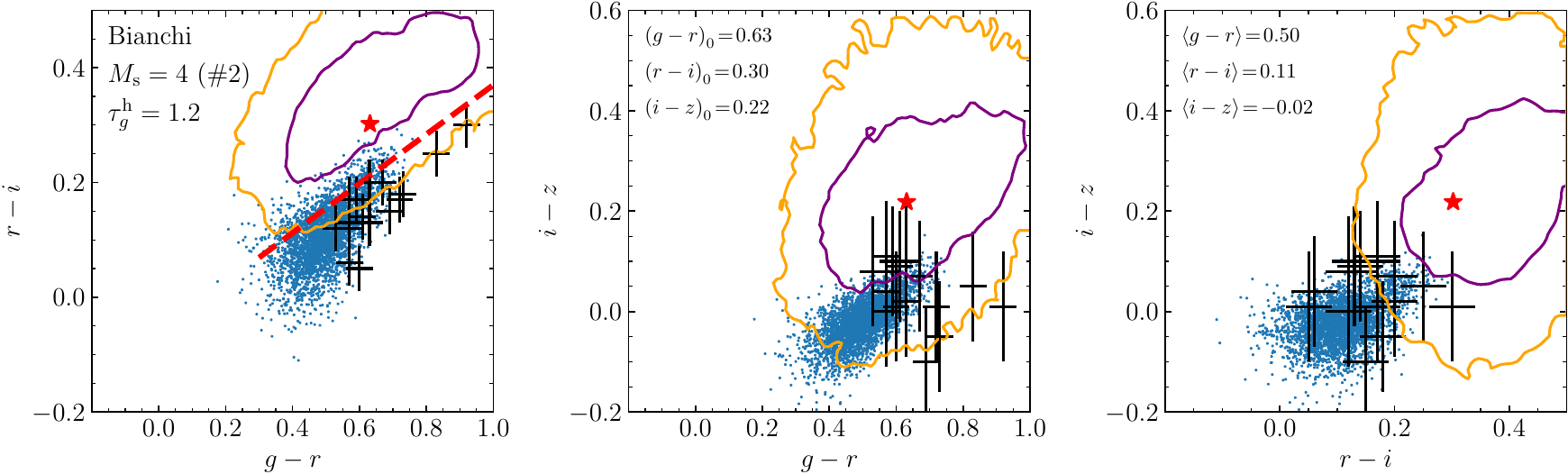}\vspace{0.3cm}
\plotone{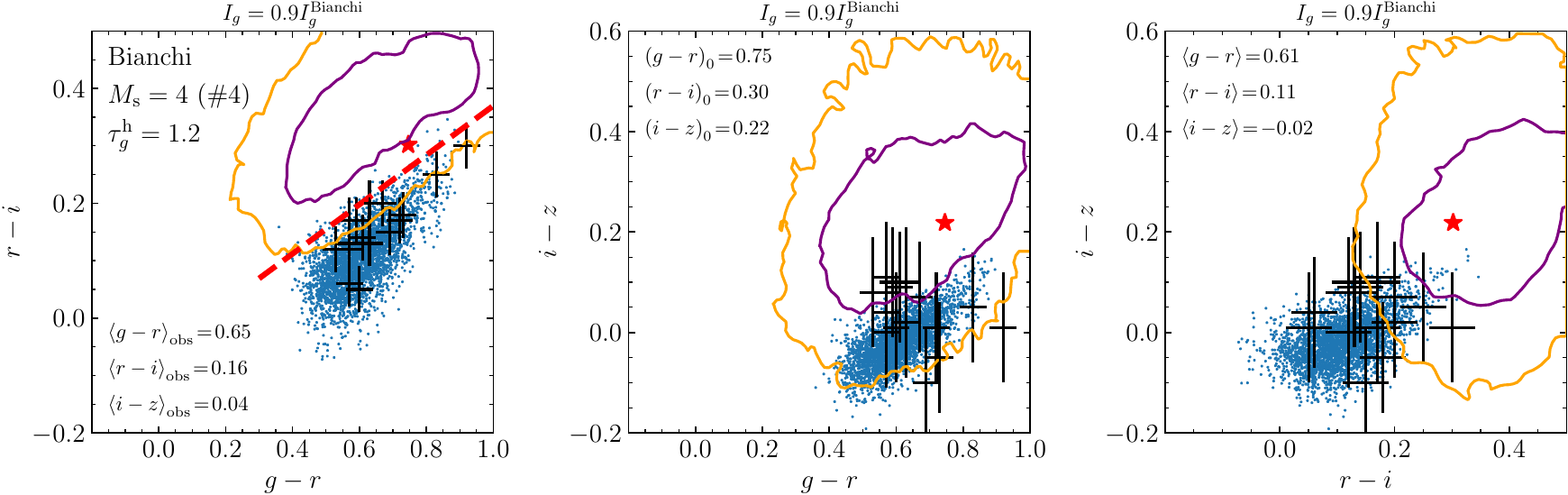}
\caption{color--color diagram for a realization with $M_{\rm s}=4$ and $\tau_g^{\rm h}=1.2$, using the ISRF of \citet{Bianchi2024}. The top and bottom rows show the color--color diagrams for the unmodified ISRF of \cite{Bianchi2024} and for the case where the $g$-band intensity is scaled down by a factor of 0.9, respectively. Star symbols indicate the initial ISRF colors. All other lines, symbols, and numerical values are the same as those in Figure \ref{fig06}.
\label{fig08}}
\end{figure*}

\begin{figure*}[ht!]
\includegraphics[width=\textwidth]{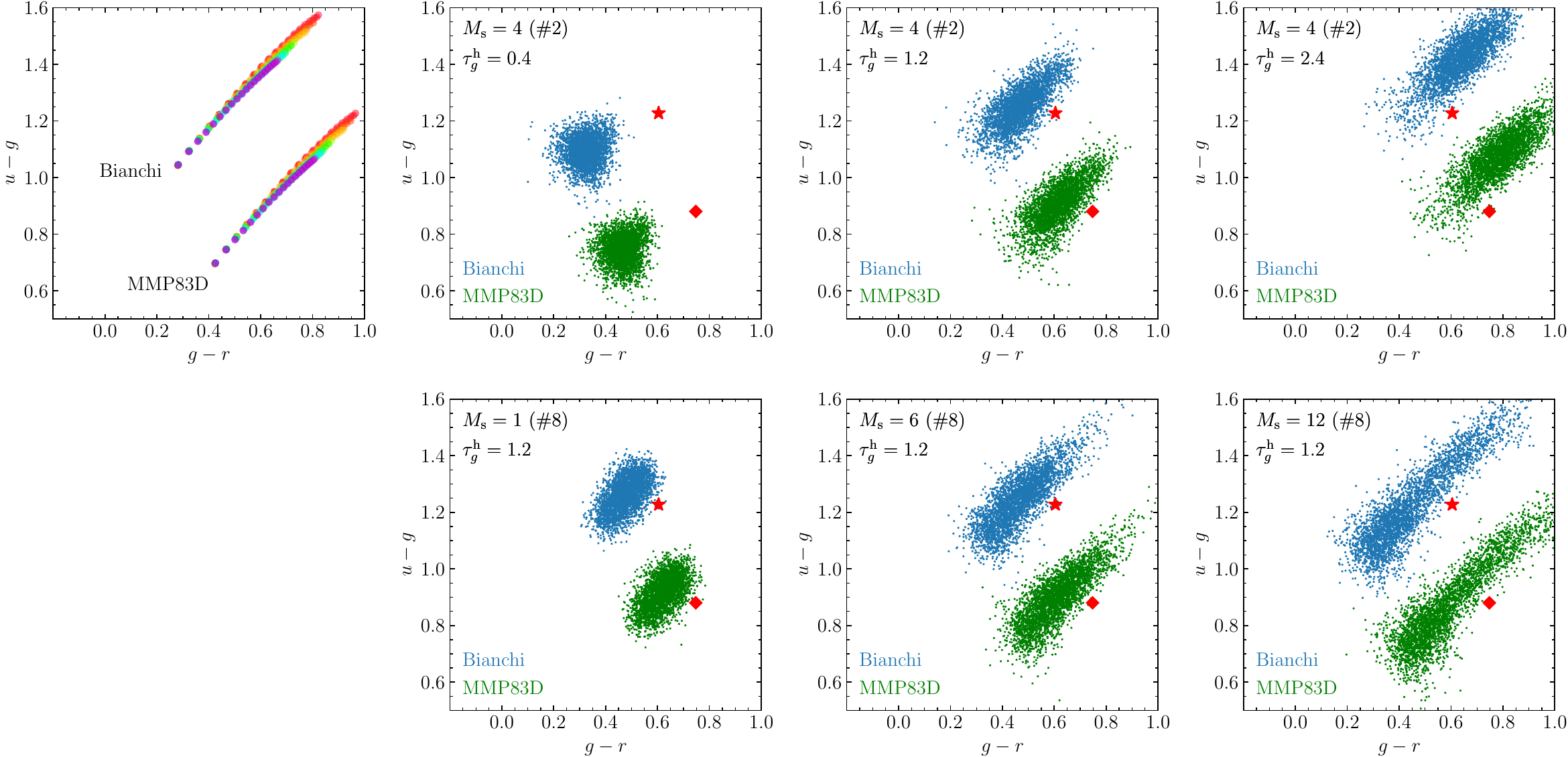}
\caption{Variation of the $g-r$ vs.\ $u-g$ color--color diagram with changing $\tau_g^{\rm h}$. The top-left panel shows the diagram constructed from the mean colors for varying $\tau_g^{\rm h}$ and $M_{\rm s}$, as in Figure \ref{fig05}. The second, third, and fourth top panels show diagrams from a typical realization (\#2) with optical depths $\tau_g^{\rm h}=0.4$, 1.2, and 2.4, respectively, at $M_{\rm s}=4$. The bottom panels show the variation of diagrams as $M_{\rm s}$ changes, with $\tau_g^{\rm h}$ fixed at 1.2. In these panels, green and blue dots represent results obtained using the ISRFs of MMP83D and \citet{Bianchi2024}, respectively. Diamond and star symbols indicate the initial input colors of the ISRFs of MMP83D and \citet{Bianchi2024}. No adjustments to the ISRFs were applied.
\label{fig09}}
\end{figure*}

\subsection{Comparison with Observations}
The purpose of this paper is not to construct a single regression-based model; rather, our goal is to identify a range of models that account for both the typical values and the dispersion in the color--color diagrams.

In Figure \ref{fig06}, the $\tau_g^{\rm h}=0.4$ model provides a relatively poor match to the three color--color diagrams, suggesting that the RT results should be shifted slightly upward and to the right.
The $\tau_g^{\rm h}=1.2$ model in the middle row provides the best agreement with the observations.
By contrast, the $\tau_g^{\rm s}=2.4$ model appears to produce slightly broader color ranges than those observed, particularly toward the redder values in $i-z$ and $r-i$.
In Figure \ref{fig07}, the model with $M_{\rm s}=1$ seems to produce somewhat smaller color ranges than those observed, whereas the model with $M_{\rm s}=12$ yields ranges that are too wide.
The models with intermediate $M_{\rm s}$ values, as shown in Figure \ref{fig06} and in the middle row of Figure \ref{fig07}, match the observed color ranges reasonably well.

In summary, by comparing the RT simulation results with the observational data, we find that the observed color--color diagrams are well explained when the Mach number is in the range $M_{\rm s} \approx 3$--6.
The observations are best, though not perfectly, matched when the optical depth of the medium is $\tau_g^{\rm h} \approx 0.8$--1.4.
However, it should be noted that this agreement is obtained only if the ISRF intensities in the $i$ and $z$ bands are reduced, or equivalently, if those in the $g$ and $r$ bands are enhanced.
Alternatively, a comparable effect can be achieved by adjusting the scattering albedos at the corresponding wavelengths.

In Figures \ref{fig05}, \ref{fig06}, and \ref{fig07}, the equation from \citet{Roman2020}, shown as red dashed lines, effectively separates the scattered light from cirrus clouds and extragalactic features, but only when the ISRF SED of MMP83D is modified.
Extragalactic features can be identified when the $r-i$ color lies above this equation for a given $g-r$ value.
It is also worth mentioning that this equation can be used to discriminate extragalactic features from cirrus clouds over a broad range of cloud optical depths, except when the depth is too high.
This is because the equation runs roughly diagonally across the diagram, similar to the track produced by increasing $\tau_g^{\rm h}$ in the color--color diagrams, as shown in Figure \ref{fig05}.
In contrast, in other color--color diagrams, the positions of cirrus clouds overlap with those of extragalactic objects at higher $\tau_g^{\rm h}$ values, because the track shifts roughly diagonally upward to the right as $\tau_g^{\rm h}$ increases.
Therefore, the $r-i$ versus $g-r$ diagram provides the best separation across a wide range of optical depths compared to other color--color diagrams.

As shown in Figure \ref{fig01}, the SED of the \citet{Bianchi2024} ISRF model has slightly lower $i$- and $z$-band intensities than those of MMP83D, suggesting that their model may be more consistent with the observed color--color diagrams.
We found that the RT simulations using the ISRF of \citet{Bianchi2024} produce similar trends for varying $M_{\rm s}$ and $\tau_g^{\rm h}$ as those for the MMP83D model, although the numerical color values differ from those obtained with MMP83D.
The upper panels of Figure \ref{fig08} compare the predicted color--color diagrams obtained using the ISRF of \citet{Bianchi2024}.
We note that the $g-r$ color appears bluer than the observed values, suggesting that the $g$-band intensity may need to be reduced.
In the lower panel of Figure \ref{fig08}, the $g$-band intensity is decreased by a factor of 0.9, and the resulting color--color diagrams are found to be consistent with the observed color--color values, similar to the case of the MMP83D ISRF model with reduced $i$- and $z$-band intensities.
Indeed, Figure \ref{fig01} shows that the ISRF of \citet{Bianchi2024}, with a reduced $g$-band intensity, closely matches that of MMP83D with reduced $i$- and $z$-band intensities.

\section{Discussion} \label{sec:discussion}

\subsection{Optical Depth and Clumpiness}
According to \citet{Roman2020}, the optical depth at 353 GHz ($\tau_{353}$) of cirrus clouds in the Stripe 82 region ranges from $\sim$$3\times10^{-6}$ to $\sim$$3\times10^{-5}$.
This optical depth can be converted to the color excess $E(B-V)$ via the empirical relation $E(B-V)\approx1.49\times 10^4\, \tau_{353}$ \citep{PlanckCollaboration2014_XI,PlanckCollaboration2016_48}.
Assuming the Milky Way dust extinction curve of \citet{CCM1989} with $R_V=3.1$, this optical depth can then be converted to that in the $g$ band: $\tau_g\approx5.08\times10^{4}\, \tau_{353}$.
From this relation, $\tau_g$ is found to lie in the range $\tau_g\approx 0.15$--1.5.
Therefore, the optical depth $\tau_g^{\rm h}$ of the dust medium inferred from the RT calculations in this paper is consistent with these values.
We also note that the linear correlations between the optical colors and $\tau_{353}$ found in \citet{Roman2020} accord well with the correlation between the mean colors and $\tau_g^{\rm h}$ shown in Figure \ref{fig04}.

Interestingly, the Spider complex has a color of $g-r \approx 0.64$, which is similar to that of the Stripe 82 region \citep{Zhang2023}.
This might not be a coincidence, as the surface brightness of scattered light in cirrus clouds is faint at low optical depths, increases with optical depth, and then becomes faint again at much higher optical depths.
In Figure \ref{fig04}, $g-r \approx 0.64$ corresponds to $\tau_g^{\rm h} \approx 1.2$, and near this optical depth the scattered light intensity reaches its maximum, as shown in Figure \ref{fig03}.
Therefore, bright cirrus clouds are likely to exhibit approximately this $g-r$ color, unless the ISRF incident on them differs significantly from that of the MMP83D model.

In this study, we assumed lognormal density media generated with a fractal algorithm to represent dust clouds, and we constrained the range of Mach number based on the color--color diagrams.
This Mach number should not be interpreted as representing the actual physical conditions; rather, it should be regarded as a parameterization of clumpiness or of the statistical properties of the density of cirrus clouds.
The variance of the log-density in a turbulent medium is approximately related to $M_{\rm s}$ according to Equation (\ref{eq01}) \citep{Federrath2008}.
Therefore, the standard deviation of $\ln \rho$ for $M_{\rm s}\approx 3$--6 is $\varsigma \approx 0.9$--1.4.
This corresponds to a standard deviation of the density, normalized by the mean density, of $\sigma/\mu \approx 1.3$--1.7.

\subsection{$u-g$ and other Colors}
Colors involving shorter wavelengths, such as $u-g$, can provide additional information for distinguishing extragalactic LSB features from cirrus clouds, as well as for constraining the SED shape of the ISRF across a wider wavelength range.
Observations at short wavelengths, or in the UV, are crucial for investigating recent star formation activity.
Although comparing models with observations in shorter-wavelength colors is beyond the scope of the present study, it is nevertheless worthwhile to present color--color diagrams that include shorter wavelengths, such as the $u$ band.

Figure \ref{fig09} shows the predicted $g-r$ versus $u-g$ color--color diagram, with the top-left panel based on the mean colors and the remaining panels illustrating diagrams from sample density realizations for different $M_{\rm s}$ and $\tau_g^{\rm h}$ values, as in Figures \ref{fig06} and \ref{fig07}.
As in Figure \ref{fig05}, the diagram forms tracks as $\tau_g^{\rm h}$ varies.
The mean colors show similar trends to those in other bands, becoming bluer as the optical depth increases and redder as the Mach number increases.
The tracks obtained with the ISRF of \citet{Bianchi2024} show redder $u-g$ colors at a given $g-r$ than those from MMP83D, as expected from the SEDs in Figure \ref{fig01}.
Even after reducing the $g$-band intensity by a factor of 0.9, the ISRF of \citet{Bianchi2024} still yields redder colors.

The second and third top panels of Figure \ref{fig09} demonstrate that the $u-g$ color of the scattered light is bluer than the original value in optically thin cases, consistent with the other colors shown in Figures \ref{fig06}--\ref{fig08}.
However, as the optical depth increases, the $u-g$ color reddens more quickly than the other colors, becoming redder than the original value at $\tau_g^{\rm h}\gtrsim 1.2$, as shown in the fourth top panel.
In contrast, the $g-r$ color begins to redden relative to its original value only when $\tau_g^{\rm h}\gtrsim 2.4$.
This rapid reddening of the $u-g$ color is also evident in Figure \ref{fig03}, where the $u$-band intensity begins to decline at $\tau_g^{\rm h}\gtrsim 1.0$, whereas the decline in the other colors occurs at higher optical depths.
Therefore, by comparing the $u-g$ color with those defined at longer wavelengths, it may be possible to constrain both the overall optical depth of the clouds and the SED of the ISRF.
The bottom panels illustrate a similar trend to that seen in the color--color diagrams of Figure \ref{fig07} as $M_{\rm s}$ increases: the mean color decreases, and the color dispersion increases, producing an extended redward tail in the distribution.

\begin{figure*}[ht!]
\plotone{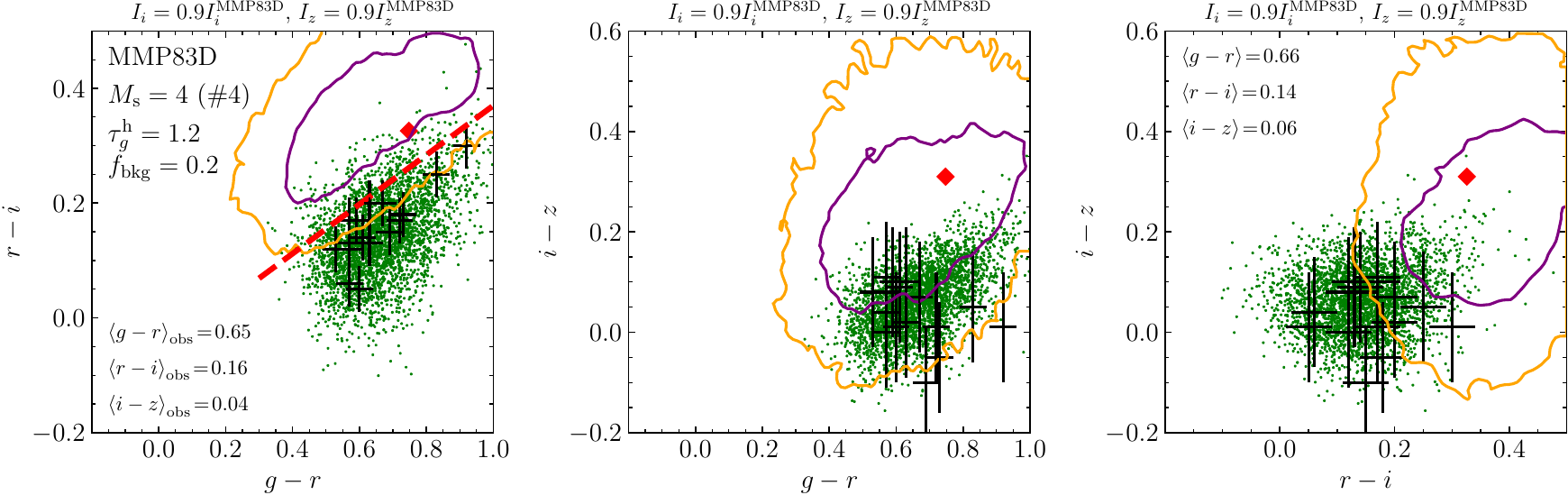}
\caption{Effect of differential photometry in color--color diagrams. Results are shown for a realization with $M_{\rm s}=4$ and $\tau_g^{\rm h}=1.2$, assuming the ISRF of MMP83D. The panels are the same as the middle row of Figure \ref{fig06}, except that differential photometry is applied using Equation (\ref{eq07}) with $\tau_0=0$. The background signal is set to $f_{\rm diff}=0.2$ of the ISRF.
\label{fig10}}
\end{figure*}

To extend coverage beyond the optical, we plan to build a next-generation space telescope that will conduct an all-sky imaging survey from the near-UV to the optical (200--700 nm) \citep{Ko2025}.
Combined with optical observations, the NUV data will place strong constraints on the SED of the ISRF and may provide an effective means of distinguishing cirrus clouds from extragalactic LSB features.

\subsection{Differential Photometry}
Observationally, the SEDs of cirrus clouds represent the differential values $I_{\rm ON}$ (nebula) minus $I_{\rm OFF}$ (sky).
As detailed in \citet{Mattila2018}, the observed surface brightness difference ($I_{\rm ON}-I_{\rm OFF}$) may be modeled using the expression:
\begin{equation}
	I_{\rm obs}(\tau)=\left[I_{\rm scatt}(\tau)-I_{\rm bkg}\times\left(1-e^{-\tau}\right)\right]e^{-\tau_0}. \label{eq07}
\end{equation}
Here, $I_{\rm scatt}$ is the scattered light of the ISRF, $I_{\rm bkg}$ is the background intensity incident on the cloud from behind, $\tau$ is the optical depth along the line of sight, and $\tau_0$ is the optical depth of the ambient dust medium in front of the cloud.
By definition, $I_{\rm bkg}\leq I_{\rm ISRF}$.
Moreover, since bright point sources in the observed images are removed in the analysis, $I_{\rm bkg}$ should be even smaller.
Therefore, we assume $I_{\rm bkg}=f_{\rm bkg}I_{\rm ISRF}$, where $f_{\rm bkg}<1$.

In Equation (\ref{eq07}), the last factor $e^{-\tau_0}$ causes the observed spectra to appear redder.
Consequently, it shifts the color--color values in the upper-right direction in the color--color diagrams, leading to an underestimation of $\tau_g^{\rm h}$ in our analysis.
However, the effect is relatively minor and does not alter our conclusion regarding the discrimination of extragalactic features from cirrus clouds.

Therefore, we examine the effect of differential photometry by ignoring $e^{-{\tau_0}}$ and allowing $f_{\rm bkg}\neq 0$, and we find that the color--color diagrams become increasingly dispersed as $f_{\rm bkg}$ increases.
Figure \ref{fig10} illustrates the case with $f_{\rm bkg}=0.2$.
The dispersion becomes excessively broad when $f_{\rm bkg} > 0.2$.
In particular, when $f_{\rm bkg}\gtrsim 0.3$, the color--color values intrude into regions occupied by extragalactic features and spread throughout the entire area in Figure \ref{fig10}, making discrimination between extragalactic features and cirrus clouds impossible.
These results cannot be attributed to random noise inherent in the Monte Carlo simulations, and the behavior is inconsistent with the observations.
For $f_{\rm bkg}<0.2$, the resulting color--color diagrams closely resemble those for $f_{\rm bkg}=0$, differing only by a broader spread as shown in Figure \ref{fig10}.
We therefore conclude that $f_{\rm bkg}$ must remain below $\sim$0.2 in the Stripe 82 region, ensuring that our present conclusions are robust.

\subsection{Gray Extinction by Large Dust Grains}
In this study, we demonstrated that a modified ISRF can bring the models and the observed colors into agreement; alternatively, modified albedos may also reproduce the observations.
Recently, observational evidence for substantial gray extinction caused by large dust grains has been reported by \citet{Siebenmorgen2025}. They found that the luminosity distance overestimates the trigonometric distance for 80\% of a sample of 33 well-known early-type stars. This discrepancy can be reconciled by incorporating a population of large, submicrometer-sized dust grains.
\citet{MRN1977} also considered the presence of such large dust grains, up to a size of 1~$\mu$m, noting that the number of large particles is not well constrained in the interstellar extinction model because they are gray.
Such large dust grains have also been employed to explain infrared reddening by \citet{Wang2015a,Wang2015b}.
The presence of large dust grains may reduce the dust albedo more strongly at longer wavelengths, such as in the $i$ and $z$ bands, thereby bringing the model colors into better agreement with the observed data.
A more detailed, quantitative analysis might be necessary; however, this is beyond the scope of the present study.

\section{Summary} \label{sec:summary}

We compared the color--color diagrams of scattered light with the observed data of cirrus clouds in the Stripe 82 region.
The scattered light is generally bluer than the incident radiation field, which in turn yields colors different from those of the stellar radiation field of our Galaxy as well as external galaxies.
This color difference provides an opportunity to distinguish diffuse features caused by Galactic dust clouds from extragalactic LSB features.
As suggested by \citet{Roman2020}, the $r-i$ versus $g-r$ diagram is the most effective among the SDSS optical color combinations for distinguishing extragalactic features from cirrus clouds.

The observed color--color diagrams provide valuable information on the spectral shape of the ISRF.
A comparison of the results obtained from the dust RT simulations with the observed data suggests that the SED of the ISRF incident upon the Stripe 82 region is lower by a factor of $\approx$0.9 in the $i$ and $z$ bands (or equivalently, higher in the $g$ and $r$ bands) than in the MMP83D model, as in the model of \citet{Bianchi2024}.
In this sense, the ISRF model of \citet{Bianchi2024} appears to better reproduce the observed colors.
However, the model of \citet{Bianchi2024} does not fully explain the $g-r$ color, which can be reconciled by reducing the intensity in the $g$ band by the same factor of $\approx$0.9.
Alternatively, the scattering albedos at the corresponding wavelengths need to be adjusted accordingly.

\begin{acknowledgments}
The work is supported by the Korea Astronomy and Space Science Institute grant funded by the Korean government (MSIT; No.\ 2025183100). KIS is supported by a National Research Foundation of Korea (NRF) grant funded by the Korean government (MSIT) (No.\ 2020R1A2C1005788).
J.L. is supported by the National Research Foundation (NRF) of Korea grant funded by the Korea government (MSIT; RS-2021-NR061998 and RS-2022-NR068800).
We are grateful to the anonymous reviewer for noting the importance of gray extinction.
\end{acknowledgments}


\bibliography{reference}{}
\bibliographystyle{aasjournalv7}


\end{document}